\documentclass[sigplan,screen,nonacm]{acmart}

\AtBeginDocument{%
  }

\usepackage{algorithm}
\usepackage{algorithmicx}
\usepackage{algpseudocode}
\usepackage{subcaption}
\usepackage{multirow}
\usepackage{balance}

\algtext*{EndIf}
\algtext*{EndFor}
\algtext*{EndWhile}
\algtext*{EndFunction}
\algtext*{EndProcedure}

\begin{document}

\title{GPU Acceleration of Sparse Fully Homomorphic Encrypted DNNs}

\author{{Lara D'Agata$^{1}$, Carlos Agulló-Domingo$^{2}$, Óscar Vera-López$^{2}$, 
Kaustubh Shivdikar$^{4}$, Ardhi~W.~B.~Yudha$^{4}$, Ferhat Yaman$^{4}$, 
David Kaeli$^{3}$, Jos\'e L. Abellán$^{2}$, Ian Colbert$^{4}$, Jos\'e Cano$^{1}$}}
\affiliation{%
\institution{University of Glasgow, UK$^{1}$ \hspace{0.3em} University of Murcia, Spain$^{2}$ \hspace{0.3em} Northeastern University, USA$^{3}$ \hspace{0.3em} AMD$^{4}$}
\country{}}

\renewcommand{\shortauthors}{D'Agata et al.}

%********************************************************************

\begin{abstract}

Fully homomorphic encryption (FHE) has recently attracted significant attention as both a cryptographic primitive and a systems challenge. Given the latest advances in accelerated computing, FHE presents a promising opportunity for progress, with applications ranging from machine learning to information security. 
We target the most computationally intensive operation in deep neural networks from a hardware perspective, matrix multiplication (matmul), and adapt it for execution on AMD GPUs. We propose a new optimized method that improves the runtime and complexity of ciphertext matmul by using FIDESlib, a recent open-source FHE library designed specifically for GPUs. By exploiting sparsity in both operands, our sparse matmul implementation outperforms its CPU counterpart by up to $3.0\times$ and reduces the time complexity from cubic to semi-linear, demonstrating an improvement over existing FHE matmul implementations.

\end{abstract}

%********************************************************************

%\begin{CCSXML}
%<ccs2012>
%   <concept>
%       <concept_id>10002978.10003022.10003028</concept_id>
%       <concept_desc>Security and privacy~Domain-specific security and privacy architectures</concept_desc>
%       <concept_significance>500</concept_significance>
%       </concept>
% </ccs2012>
%\end{CCSXML}

%\ccsdesc[500]{Security and privacy~Domain-specific security and privacy architectures}

\keywords{Fully Homomorphic Encryption, DNN Acceleration, Secure Computation, Sparse Matrix Multiplication.}

%\acmYear{2026}\copyrightyear{2026}
%\setcopyright{cc}
%\setcctype[4.0]{by-nc-sa}
%\acmConference[EuroMLSys '26]{The 6th Workshop on Machine Learning and Systems}{April 27--30, 2026}{Edinburgh, Scotland Uk}
%\acmBooktitle{The 6th Workshop on Machine Learning and Systems (EuroMLSys '26), April 27--30, 2026, Edinburgh, Scotland Uk}
%\acmDOI{10.1145/3805621.3807642}
%\acmISBN{979-8-4007-2605-7/26/04}

\maketitle

%\vspace{5cm}

\section{Introduction} 
\label{sec:introduction}

\begin{figure} [t]
    \centering
    \includegraphics[width=0.96\linewidth]{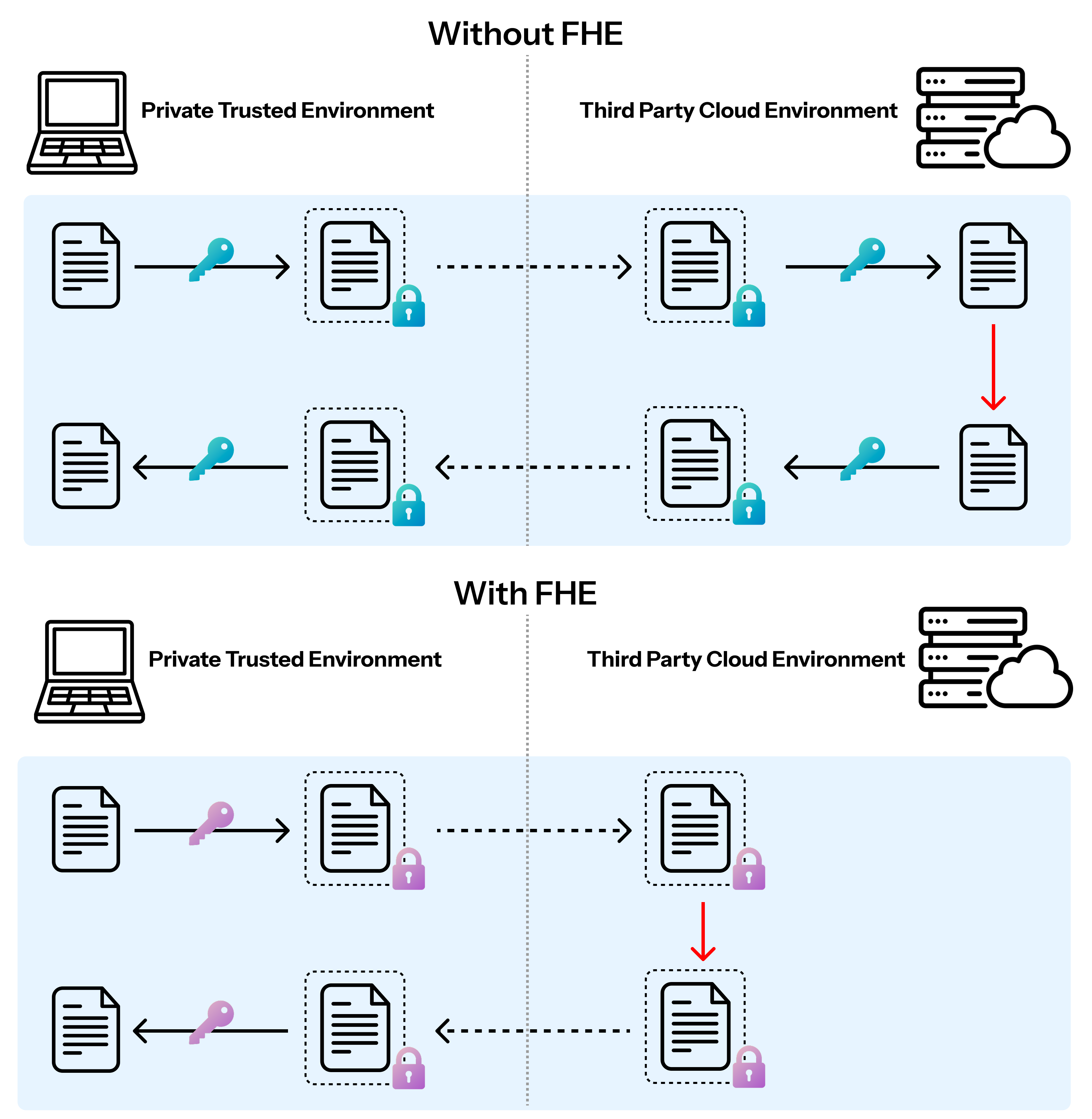}
    \caption{Visualization of a request to a third-party cloud environment with and without FHE; with FHE, data remains encrypted throughout processing.}
    \label{fig:fhe}
\end{figure}

Fully homomorphic encryption (FHE) allows arbitrary computations on encrypted data without the need for decryption, generating the same result as operations on plaintexts~\citep{Gentry2010}, and therefore holds significant promise for secure data processing in privacy-preserving computing. This process is outlined in Figure~\ref{fig:fhe}, where in a system with FHE support, a request sent over the internet is processed while still encrypted, meaning the only encryption and decryption performed are within the private trusted environment.

Being in ciphertext form, the data is difficult to operate on as it does not always conform to standard operation requirements~\citep{Vintimilla2025}; encrypted data can be much larger than plaintext data, as FHE computations involve very large polynomials. 
For instance, achieving a 128-bit security level could mean that \(n = 2^{17}\) and \(l = 35\), implying a total ciphertext size of roughly 73.4 MB. Since most hardware platforms have a cache of around 40-50 MB, the ciphertext itself would be too large to fit within the cache during an FHE operation~\citep{Agrawal2023}. 

If we also consider the key switching operation keys, with each occupying hundreds of MB, the only possibility of performing FHE operations on a modern hardware platform is through many frequent accesses to main memory. Given the large size of ciphertexts, it also follows that the computational resources it takes to perform operations on this type of data is very large in terms of both time and space. When multiplying square matrices using FHE, the runtime can be $10^6 \times$ higher compared to a multiplication with unencrypted (plaintext) data~\citep{Ferguson2025}. 

FHE schemes such as BGV~\citep{Brakerski2014} and CKKS~\citep{HEAAN} have made strides in balancing encryption complexity with computational efficiency, but despite their advancements, these schemes all face scalability challenges due to the complexity of their core operations. This computational burden poses a barrier to the widespread adoption of FHE schemes, especially in time-sensitive applications.  

Recently conducted experiments attempt to address the bottlenecks of matrix multiplication (matmul) in FHE~\citep{Ferguson2025, Ailon2025, Reddy2025, Ozcan2024, Zhou2024, Shivdikar2023}, where the problem of accelerating this process is addressed either through algorithmic improvements, such as the exploitation of sparsity, or through the creation of dedicated hardware. Sparse matrix-matrix multiplication (SpMSpM) in FHE, particularly with the CKKS scheme, faces several practical limitations. Homomorphic multiplications introduce noise that accumulates throughout the computation. Since SpMSpM involves many multiply-accumulate operations, noise can grow rapidly, potentially requiring expensive bootstrapping to maintain correctness~\citep{Ho2025}. Moreover, CKKS arithmetic introduces approximation errors, which can also accumulate across many operations~\citep{HEAAN}.

The state-of-the-art (SoTA) FHE SpMSpM approaches are limited by a few core weaknesses, including rotation and key-switching dominance, poor sparsity exploitation under packing constraints, and memory-bound execution. Even with algorithmic improvements, ciphertext rotations and rescaling dominate costs. Sparsity is difficult to leverage efficiently because SIMD (Single Instruction, Multiple Data) packing forces structured data layouts that introduce padding and redundant operations. Dedicated hardware reduces latency but is constrained by massive ciphertext sizes and memory traffic. A GPU implementation would address many of these issues, as it would benefit from massive parallelism for independent ciphertext operations, high memory bandwidth for large polynomial buffers, and efficient batched Number Theoretic Transform (NTT) acceleration, significantly reducing rotation-heavy SpMSpM latency.

In this paper, we aim to address the most computationally intensive deep neural network (DNN) operation, matmul~\citep{Gibson2025, Haris2024, Pope2022}, and devise a more efficient version of current implementations by exploiting sparsity. 
DNNs combined with sparsity have become a recent trend in FHE acceleration, particularly on GPUs, because they align well with both homomorphic computation and parallel hardware. DNN inference primarily consists of matmul operations, which map naturally to SIMD-style packed operations in schemes such as CKKS. These operations rely heavily on polynomial arithmetic and number-theoretic transforms, which GPUs can efficiently accelerate due to their support for parallelism~\citep{Wu2023}. Sparsity further reduces the number of costly ciphertext-ciphertext multiplications, thus lowering computation time, memory usage, and noise growth~\citep{Ferguson2025}. To the best of our knowledge, our new implementation represents the first FHE SpMSpM tailored for GPU usage. 
The contributions of this paper are as follows: 

\begin{itemize}
    \item A new optimized method for encrypted matrix-matrix multiplication that exploits sparsity in both operands, implemented specifically for GPUs; 

    \item A comparison of the runtime performance of the proposed approach with na\"ive and SoTA solutions on multiple high-performance AMD GPUs;

    \item A comparison of our GPU implementation against a CPU baseline built with OpenFHE~\citep{OpenFHE}, a widely used FHE library for CPUs. Our approach achieves exponential speedups over a na\"ive dense implementation and delivers approximately $2.5\times$ to $3\times$ performance improvements compared to the CPU-based baseline.
\end{itemize}

\section{Background \& Related Work}
\label{sec:background}

\subsection{CKKS}

Matrix multiplication for machine learning typically involves floating-point values and requires a large number of operations, particularly additions and multiplications, which makes efficiency and noise management critical. Given that CKKS~\citep{HEAAN} supports ciphertext packing, this dramatically reduces the number of ciphertexts and homomorphic operations required during a matmul. Although capable of packing ciphertexts, other schemes such as BFV~\citep{Fan2012} and BGV~\citep{Brakerski2014} are constrained by integer-only computation and are more susceptible to rapid noise growth and larger ciphertext sizes when dealing with real-valued or high-precision data. Moreover, CKKS is specifically designed for computations on real or complex numbers using approximate arithmetic, which is closely aligned with the needs of many machine learning, signal processing, and numerical computing applications. 

One of the distinguishing features of CKKS is its trade-off between accuracy and computational efficiency, due to its use of approximation encoding. It employs techniques such as rescaling to control noise growth~\citep{HEAAN}, ensuring computations remain feasible within the scheme's precision limits: this process adds complexity as the depth of computations increases. Unlike exact FHE schemes, CKKS does not guarantee bit-perfect correctness, but this trade-off is acceptable for many practical uses requiring real-number computations. Therefore, CKKS is especially suitable for applications such as machine learning, data analytics, and scientific computations, where inexact floating-point operations are standard.

%**************************************************************************************************

\subsection{FIDESlib}

To implement our sparsity-exploiting algorithm, we use FIDESlib~\citep{FIDESlib}, a new open-source FHE library built for working with CKKS-encrypted data on GPUs. The general software architecture of existing libraries inherently sacrifices cutting-edge performance in favor of generality, due to the many abstraction layers needed for such flexibility~\citep{FIDESlib}. 
However, FIDESlib designs software architectures tailored to specific GPU architectures to enhance performance. Most SoTA GPU-compatible CKKS libraries are not open-source~\citep{Cheddar, TensorFHE}, making FIDESlib the first openly available library built for this purpose. Moreover, we use a new HIP-compatible version of FIDESlib which makes it possible to test the algorithm on AMD GPUs.

%**************************************************************************************************

\subsection{Sparsity in FHE}

\begin{figure} [t]
    \centering
    \includegraphics[width=0.99\linewidth]{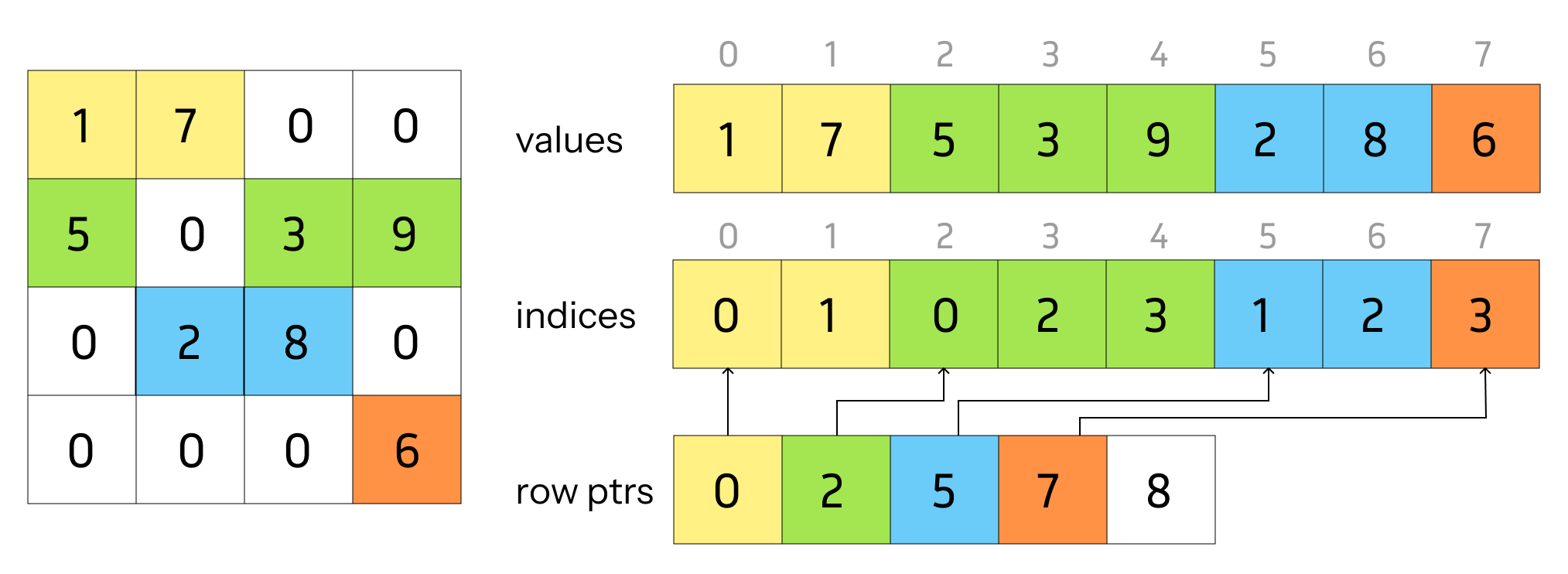}
    \caption{Example of a matrix represented in the CSR sparsity representation format.}
    \label{fig:csr}
\end{figure}

When working with large matrices, especially in high dimensional problems such as deep learning, scientific computing, and graph analytics, both memory usage and computational cost become major bottlenecks. In general, sparse processing addresses this challenge by storing and operating only on non-zero values (NZVs), thereby avoiding unnecessary arithmetic and reducing memory traffic. This improves cache utilization, reduces I/O pressure and the overall runtime compared to dense representations. Common efficient sparse formats include compressed sparse row (CSR), compressed sparse column (CSC), and block-based formats, which balance storage overhead and access efficiency depending on the access pattern. An example of CSR in shown in Figure~\ref{fig:csr}.

In the context of FHE, sparsity becomes even more important due to ciphertext expansion. After encoding and encryption in CKKS, a single plaintext value is transformed into a large polynomial ciphertext, leading to significantly increased memory and bandwidth requirements. Consequently, dense encrypted matrices quickly become impractical to store and process. To mitigate this, SIMD-style packing strategies group multiple values into a single ciphertext, enabling parallel slot-wise computation. Combining such packing techniques with sparse representations reduces the number of ciphertext-ciphertext multiplications and rotations, improving memory efficiency and making GPU-accelerated FHE matmul more scalable. By decreasing space requirements, the proposed implementation enables FHE schemes to process larger datasets, making privacy-preserving computing more scalable and cost-effective~\citep{Agrawal2023}. 

Our research is based on continuing our previous work in exploiting unstructured sparsity for matmul operations~\citep{Ferguson2025}, where we implemented a new sparsity-aware SpMSpM algorithm and compared it with other state-of-the-art FHE matmul implementations. By targeting unstructured sparsity, we were able to more accurately reflect the nature of data encountered in applications such as recommendation systems, graph processing, or natural language processing. This is particularly relevant for accelerating DNNs employing ReLU activations, which frequently induce high levels of unstructured activation sparsity~\citep{Hu2016}. We found our solution to be more efficient compared to others, being one of the first to exploit sparsity for FHE matmul. 

Other prior efforts to incorporate sparsity into FHE matmul have typically imposed notable constraints. For instance, some approaches restrict sparsity to linear systems of the form $Ax = b$, where $A$ is required to be strictly diagonally dominant~\citep{Chen2015}. 
Subsequent works attempt support for general sparse matrix multiplication~\citep{Chen2021, Cui2021}; however, they do not provide a detailed analysis of how sparsity impacts runtime performance. Moreover, these schemes are limited to dense-sparse multiplications and do not address the fully sparse-sparse setting of the operation. A recent study~\citep{Chowdhury2025} proposes a new CSR-based method for handling matrix sparsity for communications in the encrypted domain, which shows promising advancements in the acceleration of FHE operations but does not investigate the effects of sparsity on matrix multiplication.

%**************************************************************************************************

\subsection{FHE on GPUs}

A key motivation for using a GPU is the potential alignment between its architecture and the demands of FHE computation. FHE operations generally exhibit low arithmetic intensity but benefit from massive parallelism, a profile that CPUs may handle less efficiently due to their more limited thread-level parallelism and memory bandwidth. GPUs can address both constraints by exposing large numbers of compute cores and providing higher memory bandwidth to sustain the data movement these workloads require. Our library is designed to exploit both properties, with the goal of achieving throughput that would be difficult to match on a CPU.

Substantial research has explored GPU acceleration and hardware-aware optimizations for FHE. Libraries such as HEonGPU~\citep{Ozcan2024} demonstrate promising throughput improvements but lack full support for essential operations such as bootstrapping. 
Other works investigate microarchitectural GPU extensions and locality-aware scheduling to better support FHE workloads~\citep{Ailon2025, Zhou2024, Shivdikar2023}, or investigate GPU-optimized sparsity representation formats~\citep{Xu2023, Karimi2022}. 
Together, these studies highlight that, despite the progress made, there are remaining challenges in achieving efficient, scalable FHE, particularly for sparse linear algebra operations. 

By implementing our sparse FHE matrix multiplication method on top of FIDESlib, we directly address this gap by combining the throughput advantages of GPU-accelerated CKKS primitives with a sparsity-aware computation strategy, enabling scalable and practically efficient encrypted sparse matmul for GPUs.

\algrenewcommand{\algorithmicindent}{1em} 

\section{Proposed Method}

We now present our new GPU-optimized matmul implementation using a hybrid CSR (row-wise) and CSC (column-wise) representation. Unlike prior SoTA SpMSpM approaches in FHE, our method exploits complementary sparsity layouts to improve memory locality, reduce redundant rotations, and increase parallel efficiency on bandwidth-intensive accelerators. As demonstrated in Section~\ref{sec:evaluation}, this design achieves superior performance compared to existing implementations.

\textbf{Key innovation:} our method deliberately pairs row-wise and column-wise formats, enabling structured access to NZVs while minimizing ciphertext movement and rotation overhead. By ensuring that the required elements from both operands are pre-aligned within the sparse structure, we can significantly improve cache utilization and reduce key switching pressure on the GPU. 
It is important to note that our implementation performs fully encrypted ciphertext-ciphertext matmul, whereas most existing FHE SpMSpM systems assume one operand is plaintext. By supporting encrypted-encrypted multiplication, our method enables stronger privacy guarantees compared to plaintext multiplications, making our method well suited for application domains such as secure multi-party computation, privacy-preserving model training, and fully encrypted graph neural network layers, where both operands must remain confidential throughout execution.

Algorithm~\ref{alg:csr-mult} shows our method for multiplying two encrypted $N \times N$ matrices, represented in CSR and CSC format respectively. 
Storing the LHS matrix in a row-wise format and the RHS matrix column-wise offers significant performance and memory advantages, as the layout aligns well with how the matmul process works, especially for algorithms designed to operate efficiently on sparse data~\citep{Martinez2023}. It makes the row-column pairing for dot products more efficient, as the necessary data from both matrices can be accessed quickly without scanning irrelevant zeros. This format pairing also supports better memory locality and cache efficiency. Access patterns become more predictable and localized, reducing memory access overhead. It also avoids the need to transpose either matrix during computation, which is expensive for large matrices.
After being loaded at the start, all ciphertexts stay on the GPU, which we leverage by doing all server-side CKKS operations there: this transfer time is only accounted twice at runtime. Therefore, parallelism in FIDESlib is not directly exposed as it is internal to the limb partitions which the ciphertexts end up using.

\begin{algorithm}[t]
\caption{Method for multiplying two $N\times N$ matrices in CSR and CSC formats.}
\label{alg:csr-mult}
\begin{algorithmic}[1]  % Enables line numbers
\footnotesize  % Reduce font size for compactness
\For{$i = 0$ \textbf{to} N} %\Comment{N = number of rows.}
    \For{$j = 0$ \textbf{to} N} %\Comment{N = number of columns (same as rows for square matrices).}
        \State $a\_start \gets indptr_A[i], a\_end \gets indptr_A[i+1]$
        %\State $col\_under\_midpoint \gets (j < midpoint) ? 0 : 1$
        %\Comment{Establishing whether current value is before the midpoint}
        \State $b\_start \gets indptr_B[j], b\_end \gets indptr_B[j+1]$
        \State $a\_pos \gets a\_start$, $b\_pos \gets b\_start$ %\Comment{Initialize positional markers.}
        \While{$a\_pos < a\_end$ \textbf{and} $b\_pos < b\_end$} %\Comment{Iterate over NZVs in matching row x column.}
            \State $a\_col \gets indices_A[a\_pos]$, $b\_row \gets indices_B[b\_pos]$ 
            \If{$a\_col = b\_row$} %\Comment{There are matching NZVs in the current row x column pair.}
                \State $vA \gets ctxt_A$, $vB \gets ctxt_B$
                \State $min\_pos \gets a\_pos$ 
                \If{$a\_pos < b\_pos$} %\Comment{Rotate $vB$ so that the current NZV matches position in corresponding NZV in $vA$.}
                    \State $vB \gets \texttt{EvalRotate}(vB, b\_pos - a\_pos)$
                    \State $min\_pos \gets a\_pos$
                \ElsIf{$b\_pos < a\_pos$} %\Comment{Rotate $vA$ so that the current NZV matches position in corresponding NZV in $vB$.}
                    \State $vA \gets \texttt{EvalRotate}(vA, a\_pos - b\_pos)$
                    \State $min\_pos \gets b\_pos$
                \EndIf
                \State $\textbf{FHE\_SpMSpM($vA$, $vB$)}$
                \State $a\_pos++, b\_pos++$ 
            \ElsIf{$a\_col < b\_row$} %\Comment{Increment positional markers according to size of cache lines.}
                \State $a\_pos++$
            \Else
                \State $b\_pos++$
            \EndIf
        \EndWhile
    \EndFor
\EndFor
\end{algorithmic}
\end{algorithm}

%**************************************************************************************************

\subsection{SpMSpM in FHE}

\begin{figure*}
    \centering
    \includegraphics[width=0.99\linewidth]{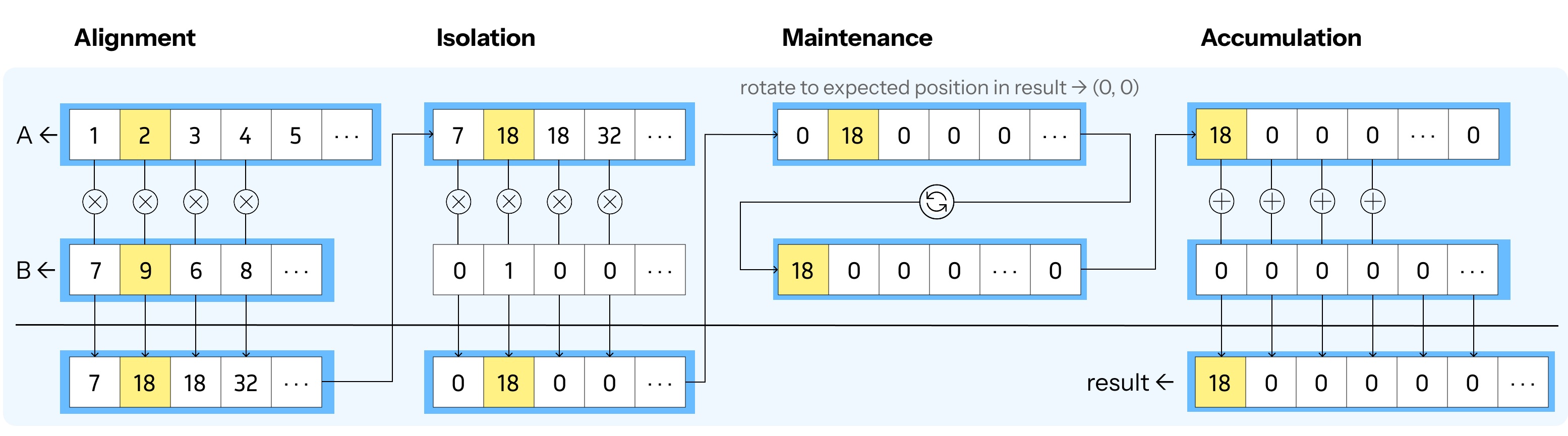}
    \caption{Overview of SIMD operations used for the FHE CKKS matmul computation, with example value matrices A and B. The encrypted vectors are represented using a blue outline, the ellipses (\texttt{...}) represent the ciphertext padding.}
    \label{fig:simd-matmul}
\end{figure*}

The SpMSpM flow is detailed in Algorithm~\ref{alg:fhe-spmspm} and visualized in Figure~\ref{fig:simd-matmul}. For each row-column pair:

\begin{itemize}
    \item \textbf{Alignment:} We identify the indices of the target NZVs and rotate one (shown in Algorithm~\ref{alg:csr-mult}). Then we perform the ciphertext-ciphertext multiplication.

    \item \textbf{Isolation:} After multiplication, we apply a plaintext mask to the product. This isolates the specific resulting value within the ciphertext.
    
    \item \textbf{Maintenance:} We perform the necessary relinearization and rescaling.
    
    \item \textbf{Accumulation:} The isolated product is rotated to its target coordinate in the result matrix and added. Because of the masking step, this addition safely updates the result matrix without corrupting existing slots.
\end{itemize}

%**************************************************************************************************

\subsection{Managing FHE constraints and overheads}

While the CSR and CSC layouts enable efficient traversal, the opacity of FHE prevents dynamic zero-skipping, which introduces a trade-off between structural privacy and performance.

\subsubsection{Metadata Transparency and Security Trade-offs}

Since the matrix values are encrypted into a ciphertext object, it is impossible to inspect its contents or execute conditional logic on it. As a result, the computation cannot dynamically skip zero entries unless some structural information about the matrix is available in plaintext~\citep{Ferguson2025}. 
To efficiently exploit sparsity, certain metadata, such as index positions, row pointers, or sparsity masks, must therefore remain unencrypted. This allows us to determine which ciphertext multiplications are necessary without decrypting the underlying values: only the sparsity pattern is exposed, while all numerical entries remain encrypted. 
For many privacy-preserving DNN inference settings, revealing structural sparsity (e.g., pruned model weights) is acceptable. However, in stricter privacy scenarios, such as when inputs use one-hot encodings, the sparsity pattern itself may leak sensitive information, since the position of nonzeros can reveal the input directly. In cases where exposing sparsity metadata is unacceptable, the server can instead rely on known plaintext model sparsity while keeping all input structure encrypted. This preserves full input privacy but limits the ability to exploit sparsity in both operands, potentially increasing inference time compared to fully sparse-aware multiplication.

\subsubsection{Mitigating overhead from intensive operations}

When dealing with ciphertexts, it is impossible to retrieve an element at a given index due to the encrypted nature of the object. Therefore, the only way to gain access to a given element within a ciphertext is to use the rotation operation, which shifts the values within the ciphertext by a given amount. This is a distinct feature of FHE computations and is very time consuming~\citep{Alves2023}, so our goal is to reduce the number of rotations as much as possible.

Algorithm~\ref{alg:csr-mult} checks the positions within the ciphertexts of the two NZVs that we want to multiply: instead of rotating both ciphertexts to the base position, only one of them is rotated to match the other. Moreover, When the position in both ciphertexts is the same, it performs no rotations and multiplies the ciphertexts as they are. In this way, we perform up to two less rotations per matching NZV pair compared to a na\"ive multiplication~\citep{Ferguson2025}. This is similar to the number of rotations in SoTA encrypted matmul methods such as Halevi-Shoup~\citep{Halevi2015} and BSGS~\citep{Wang2025}.

All encoded ciphertexts in CKKS must have a scale factor $\Delta$. The multiplication of two ciphertexts results in the product having a scaling factor of $\Delta^2$. The scaling factor must be reduced to $\Delta$ (or very close to it) before the end of the operation, or the scaling factor will eventually grow too much and allow the ciphertext modulus to overflow~\citep{Agrawal2023}. This is known as the \texttt{Rescale} operation and if not performed after every multiplication will result in decryption failure. 
After multiplying two ciphertexts, the result will be encrypted under a larger key $sk^2$ instead of the original $sk$. To ensure that the product of the multiplication will be decryptable, we require a key switch operation after each multiplication which encrypts the result back under the original secret key $sk$. This procedure is more commonly known as relinearization.

%**************************************************************************************************

\subsection{Optimized GPU formats: VCSR}

In order to investigate GPU-optimized implementations further, We implemented our method for a new sparsity representation format as well. A recent paper proposes the vertical compressed sparse row (VCSR) format~\citep{Karimi2022}, a GPU memory-aware format that focuses on achieving both high thread-level parallelism and high memory bandwidth utilization, while also minimizing compaction overhead. VCSR is designed to leverage the GPU’s coalescing unit, which can reduce global memory transactions significantly. Given that global memory transactions are the main bottleneck in SpMV, the proposed methods accelerate kernel execution significantly. VCSR also addresses format conversion overhead, an issue not addressed by earlier sparse matrix studies. 
Based on the VCSR format, we propose a new combined format by using VCSR on the LHS and VCSC on the RHS.

\begin{algorithm}[t]
\caption{FHE\_SpMSpM - Performing FHE operations for ciphertext-ciphertext multiplication.}
\label{alg:fhe-spmspm}
\begin{algorithmic}[1]  % Enables line numbers
\footnotesize  % Reduce font size for compactness
    \State $dot\_prod \gets \texttt{EvalMult}(vA, vB)$ %\Comment{Multiply the two ciphertexts.}
    \State $dot\_prod \gets \texttt{Relinearize}(dot\_prod)$ %\Comment{Relinearize to maintain low noise levels after multiplication.}
    \State $dot\_prod \gets \texttt{Rescale}(dot\_prod)$ %\Comment{Rescale.}
    \State $mask \gets \texttt{makeMask}(size^2, min\_pos)$ %\Comment{Make binary mask (only one 1 value at index of calculated dot product).}
    \State $dot\_prod \gets \texttt{EvalMult}(dot\_prod, mask)$ %\Comment{Apply mask to isolate calculated result.}
    \State $dot\_prod \gets \texttt{Relinearize}(dot\_prod)$ %\Comment{Relinearize.}
    \State $dot\_prod \gets \texttt{Rescale}(dot\_prod)$ %\Comment{Rescale.}
    %\State $final\_row\_idx \gets order_A[i] - 1$, $final\_col\_idx \gets order_B[j] - 1$
    %\Comment{Establish location of value in final matrix}
    \State $rot\_idx \gets min\_pos - (j + ( i \cdot size))$
    %\Comment{Define rotation index based on location of value in final matrix}
    \State $dot\_prod \gets \texttt{EvalRotate}(dot\_prod, rot\_idx)$ %\Comment{Match position of current dot product to the result.}
    \State $result \gets \texttt{EvalAdd}(result, dot\_prod)$ %\Comment{Add dot product to final ciphertext.}
\end{algorithmic}
\end{algorithm}

\section{Evaluation}
\label{sec:evaluation}

In order to compare our solution with baseline implementations, we tested our methods against the following: (i) \textbf{Na\"ive Dense}: it represents the baseline for our comparison, where every element in each matrix operand is multiplied individually; (ii) \textbf{Na\"ive Sparse}: a recent SoTA method~\citep{Ferguson2025} that represents a na\"ive form of sparsity exploitation, where the multiplication is still element-wise, but skips each computation where both values are zero. 
Our CSR $\times$ CSC approach is referred to as CSR/C, and our VCSR implementation as VCSR/C.
We ran tests on the four different methods, all implemented in C++ using the HIP-compatible FIDESlib library.

The methods were tested on an AMD Radeon RX 7900 XT GPU and an AMD MI300X GPU.\footnote{AMD, AMD Radeon, AMD Instinct (MI Series), and combinations thereof are trademarks of Advanced Micro Devices, Inc.} All methods were tested on the same randomly generated square matrices of sizes 8 and 16 as a proof of concept, with varying sparsity levels (0-100\%). The same benchmark tests were run five times to generate an average which captures performance variances. 
The overhead costs from converting the matrices into the appropriate format are not included: accounting for external operations such as conversion, encoding, and encryption would not allow an accurate evaluation of the performance of the matmul method itself. Moreover, our methods were also implemented in OpenFHE to include a CPU-based implementation in our results, given the lack of other GPU-based libraries built for the systems tested on.

%**************************************************************************************************

\subsection{Runtime Performance}

\begin{figure*} [t]
\centering
    \begin{subfigure}[b]{0.33\linewidth}
        \centering
        \includegraphics[width=\linewidth]{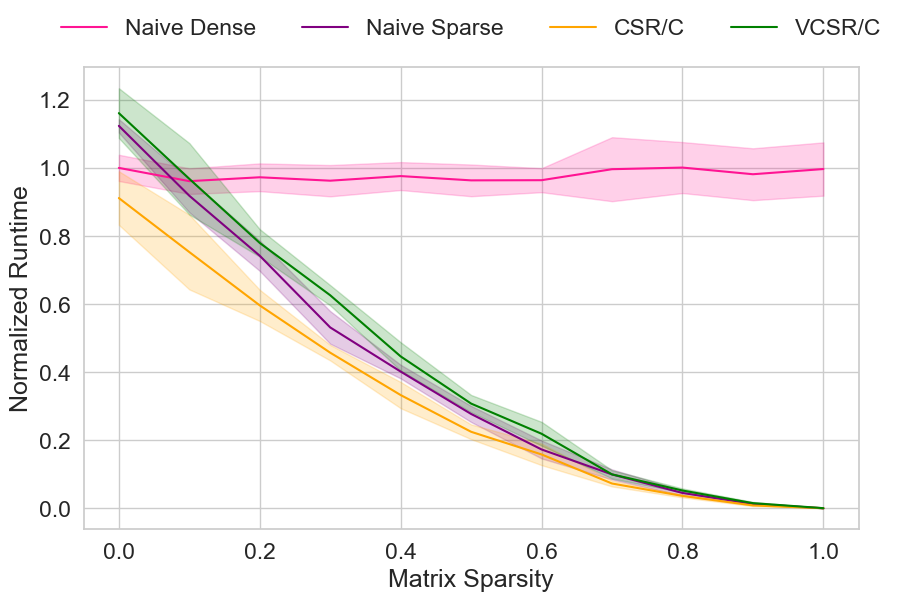}
        \caption{FIDESlib on Radeon.}
        \label{fig:runtimes_a}
      \end{subfigure}
    \begin{subfigure}[b]{0.33\linewidth}
        \centering
        \includegraphics[width=\linewidth]{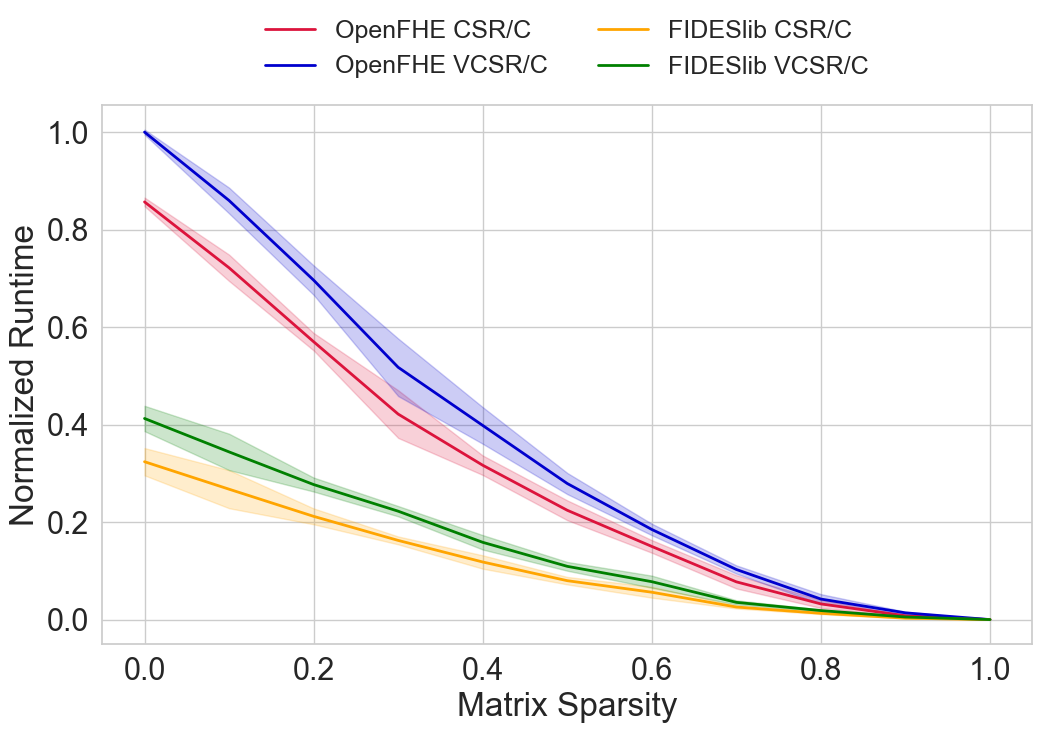}
        \caption{OpenFHE vs FIDESlib on Radeon.}
        \label{fig:runtimes_b}
    \end{subfigure}
    \begin{subfigure}[b]{0.33\linewidth}
        \centering
        \includegraphics[width=\linewidth]{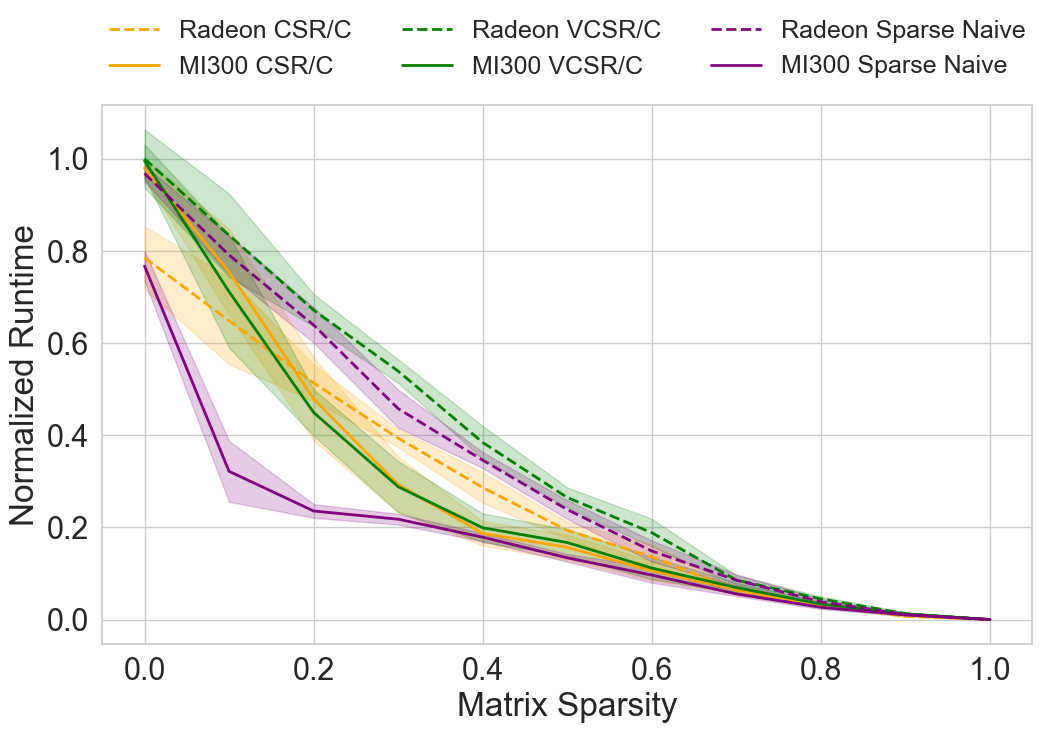}
        \caption{FIDESlib on Radeon vs MI300.}
        \label{fig:runtimes_c}
    \end{subfigure}
        
    \caption{Runtime comparison of the different algorithms used to multiply two 16x16 matrices on (a) an AMD Radeon RX 7900 GPU, (b) a comparison with OpenFHE  on the AMD Radeon GPU, and (c) runtimes of each method on the AMD Radeon and AMD MI300 GPUs. The matrix sparsity is represented as a fraction, and the runtimes are normalized.}
    \label{fig:runtimes}
\end{figure*}

\begin{table}[t]
\small
\centering
\caption{Overall time (in seconds) and speedup comparison of CSR/C and VCSR/C using FIDESlib vs OpenFHE on the AMD Radeon GPU for 16x16 matrices across sparsity levels.}
\fontsize{8.2}{9}\selectfont
\begin{tabular}{|c|c|c|c|c|c|c|}
\hline
\multirow{2}{*}{\textbf{Sparsity}} 
& \multicolumn{3}{c|}{\textbf{CSR/C}} 
& \multicolumn{3}{c|}{\textbf{VCSR/C}} \\
\cline{2-7}
& \textbf{Open} & \textbf{FIDES} & \textbf{Speed} 
& \textbf{Open} & \textbf{FIDES} & \textbf{Speed} \\
& \textbf{FHE} & \textbf{lib} & \textbf{up} 
& \textbf{FHE} & \textbf{lib} & \textbf{up} \\
\hline
0\%   & 101.234 & 38.266 & 2.65$\times$ & 118.151 & 48.743 & 2.42 $\times$ \\
\hline
10\%  & 85.261  & 31.605 & 2.70$\times$& 101.569 & 40.641 & 2.50$\times$\\
\hline
20\%  & 67.349  & 25.025 & 2.69$\times$& 82.270  & 32.730 & 2.51$\times$\\
\hline
30\%  & 49.806  & 19.190 & 2.60$\times$& 61.147  & 26.275 & 2.33$\times$\\
\hline
40\%  & 37.413  & 13.962 & 2.68$\times$& 47.057  & 18.731 & 2.51$\times$\\
\hline
50\%  & 26.497  & 9.432  & 2.81$\times$& 32.982  & 12.910 & 2.55$\times$\\
\hline
60\%  & 17.733  & 6.637  & 2.67$\times$& 21.843  & 9.188  & 2.38$\times$\\
\hline
70\%  & 9.151   & 3.052  & 3.00$\times$& 12.145  & 4.177  & 2.91$\times$\\
\hline
80\%  & 3.806   & 1.522  & 2.50$\times$& 4.978   & 2.189  & 2.27$\times$\\
\hline
90\%  & 0.929   & 0.348  & 2.67$\times$& 1.637   & 0.630  & 2.60$\times$\\
\hline
100\% & 0.002   & 0.009  & 0.23$\times$& 0.002   & 0.010  & 0.19$\times$\\
\hline
\end{tabular}
\label{tab:combined-speedup-16x16}
\end{table}

In Figure~\ref{fig:runtimes_a}, we observe the different methods running on the same GPU operating on the same matrix operands.  The na\"ive dense method has a consistent runtime independent of the sparsity level: this is expected, as the execution is independent of sparsity. The other three methods all display a near-exponential speedup compared to the na\"ive dense method, and all have consistently better runtimes starting at 10\% sparsity. We can observe that the fastest method is CSR/C, which is the only method whose runtime is faster than the na\"ive dense method even at 0\% sparsity.

Something to note in Figure~\ref{fig:runtimes_a} is that the na\"ive sparse, CSR/C, and VCSR/C methods have similar runtimes. GPUs are naturally designed for parallelism, which allows for faster, more efficient, and memory-aware matrix multiplication processes, which could be the reason for this trend. However, it would be expected for GPU-optimized sparsity formats such as VCSR to perform better than other methods when running operations on a GPU. Ciphertexts behave in complex ways, which can greatly affect the runtime and performance of a computation. Moreover, the performance of VCSR/C is highly dependent on both the sparsity pattern and the degree of sparsity~\citep{Karimi2022}. Further investigation is necessary to understand the behavior of the various methods.

In Figure~\ref{fig:runtimes_b} we can see the runtimes of the newly proposed methods implemented in OpenFHE, a non-GPU-specific library, compared to the same methods implemented in FIDES-lib. We can see that both the CSR/C and VCSR/C methods exhibit constant improvement throughout each sparsity level. Table~\ref{tab:combined-speedup-16x16} clearly displays the data, showing a constant speedup $\approx 2.7$ \( \times \) for CSR/C and  $\approx 2.5$\( \times \) for VCSR/C when the methods are run using FIDESlib compared to OpenFHE.

Figure~\ref{fig:runtimes_c} shows the runtimes of the CSR/C, VCSR/C, and sparse na\"ive implementations running on the AMD MI300X GPU. This GPU provides very large bandwidth and an even larger on-package memory, reducing stalls from ciphertext and evaluation key movement, further accelerating each method. Its architecture is optimized for sustained HPC workloads with large working sets, and its large unified memory also helps keep evaluation keys resident on-device.

%**************************************************************************************************

\subsection{Accuracy}

The accuracy of the results was calculated by comparing the resultant matrix from the CSR/C and VCSR/C methods to a baseline correct solution generated by multiplying the same matrices in plaintext form. The matrices were then compared by calculating the Frobenius norm, which is defined as the square root of the sum of the absolute squares of its elements. The difference between two matrices $A$ and $B$ can be calculated as follows: 
\begin{equation}
\small
||A - B ||_F = \sqrt{\sum_{i, j}(A_{ij} - B_{ij})^2}
\end{equation} 
As shown in Table~\ref{tab:fideslib_combined_accuracy}, both methods produce highly accurate results. Larger approximation errors are observed when multiplications are performed between larger matrices, and overall the accuracy increases with the the sparsity level. At 100\% sparsity, both matrices only have zero values, therefore no computations are happening, and the error is zero.

\begin{table}[t]
\small
\centering
\caption{Frobenius Norm of the CSR/C and VCSR/C methods comparing plaintext matmul with FIDESlib matmul across sparsity levels for matrix sizes 8x8 and 16x16.}
\fontsize{8.2}{9}\selectfont
\begin{tabular}{|c|c|c|c|c|}
\hline
\multirow{2}{*}{\textbf{Sparsity}} 
& \multicolumn{2}{c|}{\textbf{8x8 (Error e-10)}} 
& \multicolumn{2}{c|}{\textbf{16x16 (Error e-10)}} \\
\cline{2-5}
& \textbf{CSR/C} & \textbf{VCSR/C} 
& \textbf{CSR/C} & \textbf{VCSR/C} \\
\hline
0\%   & 33.54 & 38.02  & 985.58 & 1758.36 \\
\hline
10\%  & 28.47  & 23.43 & 894.49  & 526.77 \\
\hline
20\%  & 22.62  & 22.98 & 945.95  & 773.00 \\
\hline
30\%  & 22.41  & 15.44 & 591.93  & 528.83 \\
\hline
40\%  & 14.93  & 19.01 & 451.01  & 267.18 \\
\hline
50\%  & 8.79  & 10.73  & 187.34  & 213.56 \\
\hline
60\%  & 8.73  & 8.69  & 162.67  & 159.47 \\
\hline
70\%  & 3.45  & 2.94 & 51.24  & 144.36 \\
\hline
80\%  & 1.96  & 1.70 & 22.76  & 37.75 \\
\hline
90\%  & 0.34  & 0.22 & 8.57  & 9.08 \\
\hline
100\% & 0.00  & 0.00  & 0.00  & 0.00 \\
\hline
\end{tabular}
\label{tab:fideslib_combined_accuracy}
\end{table}

%**************************************************************************************************

\subsection{Time Complexity}
\label{sec:time-complexity}

The proposed optimized implementation benefits from an extremely efficient lookup 
time, due to the combination of row-wise and column-wise formats. Given that the 
algorithm iterates over all pairs of rows and columns, the number of outer 
iterations is $\mathcal{O}(N^2)$ for multiplying two $N \times N$ matrices. 
However, since the algorithm only performs computational work if the indices match, 
the actual work performed in the inner loop depends on the number of non-zero 
values (NZVs) present in each matrix. Let $k_A$ and $k_B$ denote the total number 
of NZVs in matrices $A$ and $B$, respectively. For each of the $N$ row--column 
index positions iterated in the inner loop, the number of matching operations 
scales with $\frac{k_A}{N}$ and $\frac{k_B}{N}$ (the average NZVs per row and 
column, respectively), giving an overall time complexity of:
\begin{equation}
    \mathcal{O}\!\left(N^2 \cdot \left(\frac{k_A}{N} + \frac{k_B}{N}\right)\right) 
    = \mathcal{O}\!\left(N \cdot (k_A + k_B)\right) 
    = \mathcal{O}(N \cdot k)
\end{equation}
where $k = k_A + k_B$. This expression makes the degenerate cases explicit: in 
the fully dense case, $k_A = k_B = N^2$, so $k = 2N^2$ and the complexity 
correctly recovers to $\mathcal{O}(N^3)$, matching the na\"{i}ve dense algorithm. 
Conversely, when the matrices are highly sparse and $k \ll N^2$, the complexity 
is substantially reduced. Therefore, the complexity of the optimization scales 
linearly with $N$ and $k$: this represents a large improvement compared to the 
na\"{i}ve implementations whenever the matrices are sparse. Compared to 
SoTA approaches, our implementation is similar, with time complexities 
ranging from $\mathcal{O}(\sqrt{N})$ to $\mathcal{O}(N^2)$~\citep{Halevi2015, 
Wang2025}.

%**************************************************************************************************

\subsection{Discussion}
\label{sec:discussion}

While modern DNNs that use smooth activations (e.g., GELU in Transformer models such as BERT) exhibit little intrinsic activation sparsity, practical sparsity in these workloads typically arises from weight pruning, structured sparsity, or model compression techniques, often reaching 50–90\% sparsity in practice for many deployment scenarios and production settings under typical workloads. Moreover, in the FHE setting, polynomial approximations of nonlinearities further reduce activation sparsity, making weight and operand sparsity the primary and realistic source of acceleration. In this work, we therefore focused on exploiting operand sparsity within FHE-based matmul.

\section{Conclusion}

We proposed a new GPU-optimized implementation of encrypted sparse matrix-matrix multiplication in FHE. Our implementation uses a new CSR-based sparsity representation format to reduce the complexity of the procedure, as well as a more advanced version by employing the recently proposed VCSR format. 
Our proposed matmul implementation was found to run at least 2.4$\times$ and up to 3.0$\times$ faster when implemented for GPUs. The CSR/C and VCSR/C methods are not just better for sparse matrices: the new algorithm is a generalized and efficient way to multiply packed encrypted matrices, regardless of sparsity. 
Our experiments are the first to exploit sparsity in a matrix multiplication with FHE-encrypted ciphertexts on GPUs. We show that it is possible to achieve significant speedup even when encrypting a matrix in a GPU-optimized format like VCSR/C. Although there is room for further exploration, the speedup shown in our experiments exposes the potential of running sparse-optimized FHE computations on GPUs. 

As future work, we aim to demonstrate sparsity utilization with larger matrices and extend the current approach to jointly co-design sparsity-inducing training or compression techniques tailored to FHE constraints.

\begin{acks}
This work was supported in part by Advanced Micro Devices, Inc., and the AMD AI \& HPC Cluster Program.
It was also funded through the grant CNS2023-144241 from MICIU/AEI/10.13039/501100011033 and by the European Union NextGenerationEU/PRTR. Additionally, this research was conducted within the context of the grant RYC2021-031966-I funded by MICIU/AEI/10.13039/501100011033 and by the European Union NextGenerationEU/PRTR.
\end{acks}

\newpage

\balance

\bibliographystyle{ACM-Reference-Format}
\bibliography{refs}

@misc{FIDESlib,
    title={FIDESlib: A Fully-Fledged Open-Source FHE Library for Efficient CKKS on GPUs}, 
    author={Carlos Agulló-Domingo and Óscar Vera-López and Seyda Guzelhan and Lohit Daksha and Aymane El Jerari and Kaustubh Shivdikar and Rashmi Agrawal and David Kaeli and Ajay Joshi and José L. Abellán},
    year={2025},
    eprint={2507.04775},
    archivePrefix={arXiv},   
}

@inproceedings{OpenFHE,
    author = {Al Badawi, Ahmad and Bates, Jack and Bergamaschi, Flavio and Cousins, David Bruce and Erabelli, Saroja and Genise, Nicholas and Halevi, Shai and Hunt, Hamish and Kim, Andrey and Lee, Yongwoo and Liu, Zeyu and Micciancio, Daniele and Quah, Ian and Polyakov, Yuriy and R.V., Saraswathy and Rohloff, Kurt and Saylor, Jonathan and Suponitsky, Dmitriy and Triplett, Matthew and Vaikuntanathan, Vinod and Zucca, Vincent},
    title = {OpenFHE: Open-Source Fully Homomorphic Encryption Library},
    year = {2022},
    booktitle = {10th Workshop on Encrypted Computing \& Applied Homomorphic Cryptography},
    pages = {53–63},
    numpages = {11},   
    series = {WAHC'22}
}

@InProceedings{HEAAN,
    author="Cheon, Jung Hee
    and Kim, Andrey
    and Kim, Miran
    and Song, Yongsoo",   
    title="Homomorphic Encryption for Arithmetic of Approximate Numbers",
    booktitle="Advances in Cryptology -- ASIACRYPT 2017",
    year="2017",
    pages="409--437",    
}

@misc{Cheddar,
      title={Cheddar: A swift fully homomorphic encryption library for cuda gpus},
      author={Kim, Jongmin and Choi, Wonseok and Ahn, Jung Ho},
      eprint={2407.13055},
      archivePrefix={arXiv},
      year={2024}
}

@INPROCEEDINGS{TensorFHE,
      author={Fan, Shengyu and Wang, Zhiwei and Xu, Weizhi and Hou, Rui and Meng, Dan and Zhang, Mingzhe},
      booktitle={2023 IEEE International Symposium on High-Performance Computer Architecture (HPCA)}, 
      title={TensorFHE: Achieving Practical Computation on Encrypted Data Using GPGPU}, 
      year={2023},
      volume={},
      number={},
      pages={922-934},  
}

@inproceedings{Ferguson2025,
    author = {Ferguson, Aidan and Gibson, Perry and D'Agata, Lara and McLeod, Parker and Yaman, Ferhat and Das, Amitabh and Colbert, Ian and Cano, Jos\'{e}},
    title = {Exploiting Unstructured Sparsity in Fully Homomorphic Encrypted DNNs},
    year = {2025},
    booktitle = {5th Workshop on Machine Learning and Systems},
    pages = {31–38},
    numpages = {8},  
    series = {EuroMLSys '25}
}

@article{Reddy2025,
  title={Hardware efficient arithmetic reconfigurable fully homomorphic encryption (ARFHE) accelerator of low power IoT based RISC-V processor},
  author={Reddy, T Thammi and Velagaleti, Silpakesav and Satyanarayana, BVV and Kumar, G Prasanna},
  journal={Analog Integrated Circuits and Signal Processing},
  volume={124},
  number={1},
  pages={20},
  year={2025},
  publisher={Springer}
}

@misc{Ailon2025,
      title={Changing Base Without Losing Pace: A GPU-Efficient Alternative to MatMul in DNNs}, 
      author={Nir Ailon and Akhiad Bercovich and Yahel Uffenheimer and Omri Weinstein},
      year={2025},
      eprint={2503.12211},
      archivePrefix={arXiv},     
}

@misc{Ozcan2024,
      author = {Ali Şah Özcan and Erkay Savaş},
      title = {{HEonGPU}: a {GPU}-based Fully Homomorphic Encryption Library 1.0},
      howpublished = {Cryptology Archive, 2024/1543},
      year = {2024},      
}

@inproceedings{Shivdikar2023,
    author = {Shivdikar, Kaustubh and Bao, Yuhui and Agrawal, Rashmi and Shen, Michael and Jonatan, Gilbert and Mora, Evelio and Ingare, Alexander and Livesay, Neal and Abellán, José L. and Kim, John and Joshi, Ajay and Kaeli, David},
    title = {GME: GPU-based Microarchitectural Extensions to Accelerate Homomorphic Encryption},
    year = {2023},
    booktitle = {56th Annual IEEE/ACM International Symposium on Microarchitecture},
}

@inproceedings{Zhou2024,
    author = {Zhou, Keren and Subramanian, Karthik Ganapathi and Lin, Po-Hsun and Fey, Matthias and Yin, Binqian and Li, Jiajia},
    title = {FASTEN: Fast GPU-accelerated Segmented Matrix Multiplication for Heterogenous Graph Neural Networks},
    year = {2024},
    booktitle = {38th ACM International Conference on Supercomputing},  
}

@article{Xu2023,
    author = {Xu, Weizhi and Sun, Yintai and Fan, Shengyu and Yu, Hui and Fu, Xin},
    title = {Accelerating Convolutional Neural Network by Exploiting Sparsity on GPUs},
    year = {2023},
    issue_date = {September 2023},
    volume = {20},
    number = {3},  
    journal = {ACM Trans. Archit. Code Optim.},
    month = jul,
    articleno = {36},   
}

@article{Gentry2010,
    author = {Gentry, Craig},
    title = {Computing arbitrary functions of encrypted data},
    year = {2010},
    issue_date = {March 2010},
    volume = {53},
    number = {3},
    journal = {Commun. ACM},
    month = mar,
    pages = {97–105},
    numpages = {9}
}

@INPROCEEDINGS{Chowdhury2025,
  author={Chowdhury, Moontaha Nishat and Bauer, André and Zhou, Minxuan},
  booktitle={2025 IEEE International Conference on eScience (eScience)}, 
  title={Efficient Privacy-Preserving Recommendation on Sparse Data using Fully Homomorphic Encryption}, 
  year={2025},
  volume={},
  number={},
  pages={1-9},
}

@article{Karimi2022,
    author = {Karimi, Elmira and Agostini, Nicolas Bohm and Dong, Shi and Kaeli, David},
    title = {VCSR: An Efficient GPU Memory-Aware Sparse Format},
    year = {2022},
    issue_date = {Dec. 2022},
    volume = {33},
    number = {12},
    journal = {IEEE Trans. Parallel Distrib. Syst.},
    month = dec,
    pages = {3977–3989},
    numpages = {13}
}

@inproceedings{Chen2021,
    author = {Chen, Chaochao and Zhou, Jun and Wang, Li and Wu, Xibin and Fang, Wenjing and Tan, Jin and Wang, Lei and Liu, Alex X. and Wang, Hao and Hong, Cheng},
    title = {When Homomorphic Encryption Marries Secret Sharing: Secure Large-Scale Sparse Logistic Regression and Applications in Risk Control},
    year = {2021},
    booktitle = {27th ACM SIGKDD Conference on Knowledge Discovery \& Data Mining},        
}

@ARTICLE{Chen2015,
  author={Chen, Xiaofeng and Huang, Xinyi and Li, Jin and Ma, Jianfeng and Lou, Wenjing and Wong, Duncan S.},
  journal={IEEE Transactions on Information Forensics and Security}, 
  title={New Algorithms for Secure Outsourcing of Large-Scale Systems of Linear Equations}, 
  year={2015},
  volume={10},
  number={1},
  pages={69-78},
}

@inproceedings{Cui2021,
    author = {Cui, Jinming and Chen, Chaochao and Lyu, Lingjuan and Yang, Carl and Li, Wang},
    booktitle = {Advances in Neural Information Processing Systems},
    title = {Exploiting Data Sparsity in Secure Cross-Platform Social Recommendation},  
    year = {2021}
}

@misc{Hu2016,
    title={Network Trimming: A Data-Driven Neuron Pruning Approach towards Efficient Deep Architectures}, 
    author={Hengyuan Hu and Rui Peng and Yu-Wing Tai and Chi-Keung Tang},
    year={2016},
    eprint={1607.03250},
    archivePrefix={arXiv},  
}

@article{Alves2023,
    author = {Alves, Pedro and Ortiz, Jheyne and Aranha, Diego},
    year = {2023},   
    title = {Performance of hierarchical transforms in homomorphic encryption: a case study on logistic regression inference},    
    journal = {J. of Cryptographic Engineering},
}

@misc{Haris2024,
    title={Designing Efficient LLM Accelerators for Edge Devices}, 
    author={Jude Haris and Rappy Saha and Wenhao Hu and José Cano},
    year={2024},
    eprint={2408.00462},
    archivePrefix={arXiv},  
}

@misc{Pope2022,
    title={Efficiently Scaling Transformer Inference}, 
    author={Reiner Pope and Sholto Douglas and Aakanksha Chowdhery and Jacob Devlin and James Bradbury and Anselm Levskaya and Jonathan Heek and Kefan Xiao and Shivani Agrawal and Jeff Dean},
    year={2022},
    eprint={2211.05102},
    archivePrefix={arXiv},    
}

@article{Brakerski2014,
    author = {Brakerski, Zvika and Gentry, Craig and Vaikuntanathan, Vinod},
    title = {(Leveled) Fully Homomorphic Encryption without Bootstrapping},
    year = {2014},
    issue_date = {July 2014},
    volume = {6},
    number = {3},
    journal = {ACM Trans. Comput. Theory},
    month = jul,
    articleno = {13},
    numpages = {36}
}

@misc{Fan2012,
      author = {Junfeng Fan and Frederik Vercauteren},
      title = {Somewhat Practical Fully Homomorphic Encryption},
      howpublished = {Cryptology Archive, 2012/144},
      year = {2012},
}

@book{Agrawal2023,
    title={On Architecting Fully Homomorphic Encryption-based Computing Systems},
    author={Agrawal, Rashmi and Joshi, Ajay},
    year={2023},
    publisher={Springer Cham},
}

@InProceedings{Vintimilla2025,
    author="Vintimilla-Tapia, Emmanuel
    and Rojas, Alexander
    and Sig{\"u}enza, Marco
    and Rodr{\'i}guez Z{\'u}{\~{n}}iga, Andrea Paulina
    and Cedillo, Priscila",    
    title="A Systematic Literature Review on the Security Weaknesses of Fully Homomorphic Encryption Schemes",
    booktitle="Information and Communication Technologies",
    year="2026",
    publisher="Springer Nature Switzerland",
    address="Cham",
    pages="312--328"
}

@inproceedings{Ho2025,
    author = {Ho, Ming-Chien and Ku, Yu-Te and Xiao, Yu and Liu, Feng-Hao and Hsu, Chih-Fan and Chang, Ming-Ching and Hung, Shih-Hao and Chen, Wei-Chao},
    title = {Invited Paper: Efficient Design of FHEW/TFHE Bootstrapping Implementation with Scalable Parameters},
    year = {2025},      
    booktitle = {43rd IEEE/ACM International Conference on Computer-Aided Design},    
}

@inproceedings{Wu2023,
    author = {Wu, Yannan Nellie and Tsai, Po-An and Muralidharan, Saurav and Parashar, Angshuman and Sze, Vivienne and Emer, Joel},
    title = {HighLight: Efficient and Flexible DNN Acceleration with Hierarchical Structured Sparsity},
    year = {2023},
    booktitle = {56th Annual IEEE/ACM International Symposium on Microarchitecture},
}

@inproceedings{Martinez2023,
    author = {Mu\~{n}oz-Mart\'{\i}nez, Francisco and Garg, Raveesh and Pellauer, Michael and Abell\'{a}n, Jos\'{e} L. and Acacio, Manuel E. and Krishna, Tushar},
    title = {Flexagon: A Multi-dataflow Sparse-Sparse Matrix Multiplication Accelerator for Efficient DNN Processing},
    year = {2023},   
    booktitle = {28th ACM Int. Conference on Architectural Support for Programming Languages and Operating Systems},
}

@InProceedings{Halevi2015,
    author="Halevi, Shai
    and Shoup, Victor",  
    title="Bootstrapping for HElib",
    booktitle="Advances in Cryptology -- EUROCRYPT 2015",
    year="2015",
    publisher="Springer Berlin Heidelberg",    
}

@inproceedings{Wang2025,
    author = {Wang, Qingfeng and Wang, Li-Ping},
    title = {A Novel Asymmetric BSGS Polynomial Evaluation Algorithm under Homomorphic Encryption},
    year = {2025},  
    booktitle = {20th ACM Asia Conference on Computer and Communications Security},    
}

@article{Gibson2025,
    title = {{{DLAS}}: {{A Conceptual Model}} for {{Across-Stack Deep Learning Acceleration}}},
    shorttitle = {{{DLAS}}},
    author = {Gibson, Perry and Cano, Jose and Crowley, Elliot and Storkey, Amos and O’boyle, Michael},
    year = {2025},
    journal = {ACM Trans. Archit. Code Optim.}
}

\end{document}